\documentclass[apj]{emulateapj}

\begin{document}
\bibliographystyle{apj}
\title{The Flow Around a Cosmic String, Part I: Hydrodynamic Solution}
\author{Andrey Beresnyak}
\affil{Los Alamos National Laboratory, Los Alamos, NM, 87545}
\affil{Nordita, KTH Royal Institute of Technology and Stockholm University, SE-10691}

\begin{abstract}
 Cosmic strings are linear topological defects which are hypothesized to be produced during inflation.
 Most searches for strings have been relying on the string's lensing of background galaxies or CMB.
 In this paper I derive a solution for the supersonic flow of the collisional gas
 past the cosmic string which has two planar
 shocks with shock compression ratio that depend on the angle defect of the string and its speed.
 The shocks result in compression and heating of the gas and, given favorable condition, particle acceleration.
 The gas heating and overdensity in an unusual wedge shape can be detected by observing HI line at high redshifts.
 The particle acceleration can occur in present-day Universe when the string crosses the hot gas contained
 in galaxy clusters and, since the consequences of such collision persist for cosmological timescales, could
 be located by looking at the unusual large-scale radio sources situated on a single spatial plane.
\end{abstract}

\keywords{cosmology: theory---hydrodynamics---shock waves: acceleration of particles---radio continuum: general}

\section{Introduction} 
Cosmic strings are hypothetical objects generically predicted by most modern inflationary models and are expected to survive
till present time as large over-horizon kinked linear objects and smaller loops with about 10 horizon-scale strings in the observable
portion of the Universe \citep{Polchinski2007}. The velocities of strings and loops are
expected to be trans-relativistic such with the rms velocity $v_s\approx 0.7c$. The main parameter
of the string is the symmetry breaking scale $\eta$ which determines the mass per unit length $\mu=\eta^2$ and
the angle defect of the straight string $\theta=8\pi G \mu/c^2$. For a general introduction of cosmic strings see,
e.g., \citet{vilenkin1994, hindmarsh1995,copeland2011}.
Cosmic strings are expected to lens background sources of light \citep{vilenkin1984,morganson2010,sazhina2011}
and cosmic microwave background \citep{vilenkin1986}.
The loops and kinks will emit gravitational waves \citep{spergel1987}
and leave wakes behind themselves \citep{sornborger1997,duplessis2013}. This will affect structure formation
in the early Universe \citep{khatri2008,shlaer2012}.
Kinks tend to straighten themselves by emitting gravitational waves and loops tend to evaporate for the same reason.

The current upper limits on the angle defect $\theta$ come from the lensing of background galaxies, $\theta<6\times 10^{-5}$ \citep{morganson2010},
and the CMB lensing, $\theta<7\times 10^{-6}$ \citep{wyman2005}. The pulsar timing experiments \citep{Damour2005} give tighter limits,
but are more model-dependent.
It is potentially interesting to look for direct interaction of strings with ordinary collisional matter. This subject was mostly overlooked in the literature.
The straight segment of the string could produce interaction signatures that are peculiar in that they lay on a single spatial plane. 

The paper is organized as follows. Section~2 describes an exact hydrodynamic solution of a homogeneous flow past a linear angle defect. Section~3 estimates over-densities and heating produced by the shocks behind the string
in the early Universe and briefly discusses observational possibilities compared with previously reported dark matter
wakes. Section~4 estimates particle acceleration on the shocks and their potential observability.
Section~5 contains the discussion.

\section{Supersonic flow past the linear angle defect}
Ordinary matter in a form of either neutral or ionized gas is normally considered within a fluid framework
due to its relatively high collisionality. In the case of atomic gas the mean free path in hydrogen right after recombination is around $4\times10^{-5}$pc, which is tiny compared with cosmological scales or the scales of the string.
The ordinary matter should, therefore, be considered collisional for the purpose of considering a large-scale solution
of a flow around the string. 

Below I will describe a supersonic flow of ordinary collisional fluid past the string (for the estimate of collisionality see Section~3). The straight string segment has no gravity of its own, but is manifested by the presence
of an angle defect. It is convenient to consider the flow in the string rest frame and to map the space around it onto the Euclidean space.
As the perpendicular cross-section of the string is a cone, the projection involves an angular cut in flat space with the sides of the cut mapped
onto each other.
Fig.~1 shows perpendicular cross-section of the space around the string, where I have chosen to use the cut with sides, which are parallel
to the fluid velocity. Such a cut ensures that the flow pattern is symmetric with respect to the cut direction.
The flow changes its velocity from ${\bf v_1}$ to ${\bf v_2}$ at two oblique shocks having angle $\beta$ with the tail direction $x$.
I will also designate the deflection angle $\alpha=\beta+\theta/2$ and the ratio of specific heats $\gamma$ and I assume $\gamma=5/3$,
as I mostly deal with either monoatomic gas or very cold hydrogen. I will also introduce the $\beta_s=v_1/c$ and the Mach number of the inflow 
$M_1=v_1/c_s$, where $c_s$ is the sound speed. For electron-proton plasma $M_1$ can be approximated by
$1.68\times 10^4 \beta_s (T/1{\rm eV})^{-1/2}$, where $T$ is the plasma temperature.

Applying conservation of matter, momentum and energy to the flow depicted on Fig.~1, and excluding most variables, 
I arrive at the oblique shock relation see, e.g., \citet{ll06}, where the deflection angle and the shock
angle are related by the angle defect of the string:

\begin{equation} 
\cot \frac{\theta}{2}=\tan \alpha \left [ \frac{(\gamma+1)M_1^2}{2(M_1^2 \sin^2\alpha-1)}-1\right ], \label{oshock}
\end{equation}

The Equation~\ref{oshock} can be solved for $\alpha$ and has two branches of solutions, one of which, giving larger
$\alpha$ is only realized in confined geometries, while in open boundary flows the solution with smaller $\alpha$
is realized \citep{ll06}. The expected values for the angle defect $\theta$ for astrophysical strings are fairly small, $<10^{-4}$, see \S 1, and on the second branch of the solution of Equation~\ref{oshock} this means that $\alpha, \beta \ll 1$ as well. Assuming that $\alpha,\beta, \theta \ll 1$, but $M_1 \theta$, $M_1 \beta$ are not necessarily small,
the second branch will give the following equation for $\alpha$:

\begin{equation}
4M_1^2\alpha^2-\alpha M_1^2\theta(\gamma+1)-4=0. \label{oshock2}
\end{equation}

If $\theta \gg 1/M_1 \approx 10^{-4}\beta_s^{-1}T^{1/2}$, the shocks are strong and solving (\ref{oshock2}) gives $\beta=\theta(\gamma-1)/4$.
In the opposite limit, $\theta \ll 10^{-4}\beta_s^{-1}T^{1/2}$ the shock is weak and the solution
is $\beta=1/M_1$, realizing the ``Mach cone''. In the case of  weak shock the effective Mach number in the frame
of the shock is $M=1+\theta(\gamma+1)M_1/8$ and the compression ratio
is $\rho_2/\rho_1=1+\theta M_1/2$, while in the case of a strong shock it is $M=\theta(\gamma+1)M_1/4$ and the compression
ratio approaches $(\gamma+1)/(\gamma-1)$.

It is potentially interesting to look for the flow solutions in magnetized media, i.e. to consider the magnetohydrodynamic (MHD)
problem. The general orientation of the field will break the symmetry of the flow that I have used to derive its structure
on Fig.~1, so such a flow will be more complex. Two special MHD cases can be treated relatively easily, however.
If the magnetic field is parallel to the string, both shocks will be perpendicular shocks and the shock condition is the same,
except for adding magnetic pressure to the plasma pressure \citep{ll08}. If the magnetic field is perpendicular to both the string and the inflow
velocity and $\theta$ is small, the solution will be similar to hydrodynamic solution.
The general MHD case will be considered elsewhere.

\section{Detection in the early Universe by 21 cm line}
As I demonstrated in Section~2, strings will leave behind a wake of compressed and heated material, which has a well-defined
shape, over-density and dimensions, depending on the Mach number of the flow and the angle defect of the string. So far, cosmic string wakes
has been considered primarily in the collisionless medium where they produce the wedge-shaped wake with angle $\theta/2$ and over-density of two,
see, e.g. \citep{silk1984,brandenberger2010,tashiro2013,brandenberger2014}. The gas will subsequently
accrete on the dark matter to produce features in HI. This subsequent will happen on much later times, however.
In this Section I will neglect the self-gravitational effect of the wake and concentrate on the direct
hydrodynamic interaction between the gas and the string.

I will assume that $M\gg 1$, $\gamma=5/3$, which should be the case for molecular hydrogen with $T<70$K. 
The over-density and temperature of such supersonic trail right after the interaction with the string
could be expressed as:

\begin{equation}
\rho_2/\rho_1=4/(1+3M^{-2})\approx 4,
\end{equation}

\begin{equation}
T_2/T_1\approx \frac{5}{16} M^2.
\end{equation}

Assuming that the hydrogen temperature scales adiabatically as $(1+z)^{3(\gamma-1)}=(1+z)^2$ after $z=500$, the effective Mach number will be

\begin{equation}
 M=\frac{2}{3}\theta M_1=120\left(\frac{\theta}{10^{-5}}\right)\left(\frac{\beta_s}{0.5}\right)\left(\frac{1}{1+z}\right),
\end{equation}

that is, I expect the shocks from the string to become strong starting from $(1+z)\approx 120$ and at later times before the re-ionization,
and produce heating and compression resulting in the excess of 21 cm emission due to both higher temperature and high density.

It should be noted that equations (3-5) describe overdensity and heating only as a function of redshift and the angle defect. This makes them distinctly different from expressions obtained for gravitational accretion on wakes which
has an extra unknown, which is the time allowed for accretion. Our expression will, therefore, be easier to use for
direct estimation of $\theta$, given the redshift and the excess 21cm emission, however special care should be given
for not confusing the two effects.

\begin{figure}
\begin{center}
\includegraphics[width=0.4\columnwidth]{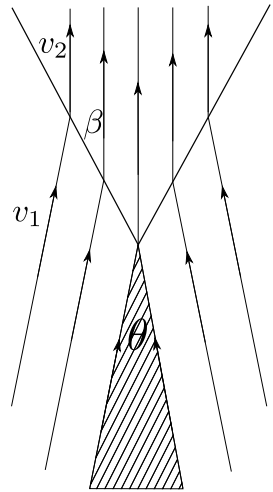}
\end{center}
\caption{Flow around a linear topological angle defect of $\theta$. I am using Euclidean 2D plane with a dashed area representing an angular cut $theta$,
which is parallel to the flow velocity.}
\label{shocks}
\end{figure}

\section{Detection by radio emission}

Shocks propagating through magnetized plasma tend to accelerate particles and produce radio emission
by synchrotron mechanism and $\gamma$-ray emission through inverse Compton mechanisms. I will consider string propagating through the present-day well-ionized
intergalactic or intracluster medium and estimate the effects of shock acceleration.  
Using the effective Mach number for oblique shocks derived in Section~2, the change 
in enthalpy of the gas per unit time per unit area -- the power, in principle available for acceleration,
on both shocks can be estimated as 

\begin{eqnarray}
P_s=3n m_p c_s^2 c \beta_s \theta (\gamma+1)/4=\nonumber \\
3.2\times 10^{-6} {\rm \frac{erg}{{\rm cm}^2 s}} \frac{n ({\rm cm}^{-3})}{10^{-3} }
\frac{T}{1\rm keV} \beta_s \frac{\theta}{10^{-5}}, 
\end{eqnarray}

for $\theta \ll 10^{-4}\beta_s^{-1}T^{1/2}$, or

\begin{eqnarray}
P_s=n m_p c^3 \beta_s^3 \theta^3 ((\gamma+1)/4)^3=\nonumber \\
1.3\times 10^{-5} {\rm \frac{erg}{{\rm cm}^2 s}} \frac{n ({\rm cm}^{-3})}{10^{-3} }
\beta_s^3 \left(\frac{\theta}{10^{-3}}\right)^3,
\end{eqnarray}

for $\theta \gg 10^{-4}\beta_s^{-1}T^{1/2}$.

Radiation efficiencies of the shocks are fairly uncertain, for several reasons. The first-principle
calculation of acceleration efficiencies are still not available due to the complex nature of the acceleration
process \citep{Malkov2001}, however some results based on phenomenological model for particle scattering is
available \citep{kang2012}. For the same reason, the injection process is not fully understood and the
electron/proton ratio is not exactly known. While in supernova shocks, given typical densities and shock
Mach numbers, the amplified magnetic field at the shock dominates over radiation field and synchrotron losses dominate over inverse Compton losses, in the tenuous ICM and IGM, the opposite could be true. The conventional approach to deal with these uncertainties is to introduce parameters such as acceleration efficiency
and the magnetic field amplification efficiency and make educated guess based on available theory
studies as well as observations, see, e.g. \citet{keshet2004}. I will further simplify the above approach, introducing
the radio emission efficiency of $\eta_r$, keeping in mind that it depends on the gas density and the Mach number
in a fairly complex way. We would expect radiation efficiency in the range $10^{-2}-10^{-6}$.
The radio spectrum is fairly uncertain for the same reasons as above. I can estimate the spectral
brightness near the peak of the emission using the total emitted power
as $\nu_m I_\nu(\nu_m) \approx P_s \eta_r / 4 \pi \sin i$, where
$i$ is an angle between the line of sight and the velocity of the string, provided that it is not much smaller than $1/M_1$. I obtain the following expression for the surface brightness temperature $T_b=I_\nu c^2/2k\nu^2$:

\begin{eqnarray} 
T_b=83{\rm mK}(\sin i)^{-1} \frac{\eta_r}{10^{-4}}
\left(\frac{\nu_m}{1 \rm GHz} \right)^{-3} \nonumber \\
\times\frac{n}{10^{-3} {\rm cm}^{-3}}
\frac{T}{1\rm keV}\beta_s \frac{\theta}{10^{-5}},
\end{eqnarray}

where I used the weak shock case. 

The new generation of radio telescopes, such as SKA, should be able to detect such low surface brightness objects, e.g., 50\% of the SKA will detect 5 mK surface brightness at the $3\sigma$ level
with beam of 8'' in 1 hour \citep{Feretti2004}, however the problem of confusion with
other objects and dealing galactic background is indeed quite challenging. Several morphological
features specific to the remnants of string activity can help in differentiating these objects, however.
Speaking of the collision of the string with the galaxy cluster, other large-scale ($>$1Mpc)
low surface brightness objects expected to be detected with the new generation of radio telescopes
include intergalactic shocks \citet{keshet2004,keshet2004b}, accretion shocks, currently detected
only in some clusters as the so-called radio relics \citep{weeren2010}, and diffuse radio halos \citep{carilli2002,cassano2010}, currently detected only in merging clusters but thought to be
present universally. Out of these three, all have different morphological and/or radiative
features compared to remnants of string activity. Intergalactic and accretion shocks are expected to be 
detected on the outskirts of the cluster, where the surface brightness is being enhanced due to the
projection effect. For the string shocks, the projection factor $(\sin i)^{-1}$ is, basically, a constant,
while the surface brightness should strongly increase with higher density, towards the center of 
the cluster. Both types of objects are expected to emit significantly polarized radio emission.
Comparing cluster halos and string trails, the former are unpolarized and rather spherical, mimicking
the shape of the cluster, while the latter will be polarized and, in general, rather elliptic,
as they cross the cluster at some angle with respect to the field of view. In fact, due to the selection
effect, the trails with higher projection factor will be much more likely observed, while their morphology
will be the most unusual -- basically a thin bright stripe across the cluster.  

Finally, the surface of past interaction of the string with dense matter over a Hubble time will be very large\footnote{Also, cluster sound crossing times are cosmological, while for bigger objects, such as filaments, sound crossing times are much larger than the age of the Universe. I also can ignore the shock propagation
speeds as far as the coincidence detection method described below is concerned.} and it is likely to have patches where the shocks will be amplified, when propagating down the density gradients \citep{ostriker1988}.
Also, the acceleration efficiency of the twin shock is higher than that of a single shock, as the downstream particle traveling from one shock could diffuse to the upstream of the other \citep{melrose1993}.

Another method to differentiate string trails and other objects can rely on the strings themselves
and their trajectories being fairly straight on sub-horizon scales. This means 
that the straight segments  of the string will leave behind relics which lay on the single spatial plane.
Searching for spatial planes that contain significant number of
large-scale ($>$1Mpc) radio sources could be another viable method which will allow
to avoid confusion with accretion/intergalactic shocks.

\section{Discussion}
In this paper I presented the solution of the flow of collisional matter around the cosmic string for
the first time. Aside from shocks in collisional medium,
strings can also produce wakes in dark matter \citep{silk1984,brandenberger2010}, which also has
wedge shape, but with a constant angle of $\theta/2$ and the constant dark matter compression ratio of two, independent on $\theta$. The subsequent self-gravitational contraction
of such wakes will also draw in ordinary matter, possibly resulting in secondary shocks \citep{hernandez2012} and various observational effect of the entrained hydrogen, e.g. the enhanced 21 cm emission \citep{tashiro2013,brandenberger2013,brandenberger2014}. As the wake gravitationally contracts, entrains and heats hydrogen, it no longer presents such a clear and well-defined angular shape.
In contrast, the trail in collisional matter described in this paper heats the gas momentarily.
The relative importance of the trails considered here and the trails produced by collapse
of dark matter wakes to the HI structures in the early Universe
will be considered in a future publication.

Given that the typical string segment length as well as their distance between each other is of order of 1 Gpc, the radio searches for strings should survey large-scale distant objects, such as clusters, and focus on extended
sources at least 1 Mpc in physical size. The cluster radio halos could be confused with the string trails,
but they are associated with turbulent acceleration in clusters \citep[see, e.g.,][]{brunetti2007,BXLS13} that underwent recent merger, so surveying smaller and quieter clusters is more advantageous. Also, as I pointed out above should have different polarization properties and morphology. The collision of the string with giant molecular clouds (GMC) could in principle produce much stronger signal, e.g. for $\theta=10^{-5}$, $T=10K$ the shocks will
have an effective Mach number of 3, and assuming density of $10^3 {\rm cm}^{-3}$, $\beta_s=0.5$ and the
acceleration efficiency of $10^{-2}$, the
surface brightness temperature is around 4 K. Given the small volume fraction of GMCs in the Universe, such collision is fairly unlikely, however.

\section{Acknowledgments}
I am grateful to Avi Loeb for a discussion. This work was supported by the LANL/LDRD program and the DoE/Office of Fusion Energy Sciences. I am grateful for the hospitality of Nordita. 

\

\def\apj{{\rm ApJ}}           
\def\apjl{{\rm ApJ }}          
\def\apjs{{\rm ApJ }}          
\def\grl{{\rm GRL }}
\def\aap{{\rm A\&A } }
\def\mnras{{\rm MNRAS } }
\def\physrep{{\rm Phys. Rep. } }               
\def\prl{{\rm Phys. Rev. Lett.}} 
\def\jcap{{\rm JCAP}}
\def\pre{{\rm Phys. Rev. E}} 
\def\nar{{\rm New Astronomy Reviews}}

\bibliography{all}

\end{document}